\documentclass[10pt,a4paper]{article}
\usepackage[left=2cm,right=2cm,top=2cm,bottom=2cm]{geometry}
\usepackage[english]{babel}
\usepackage{amsmath}
\usepackage{amsfonts}
\usepackage{amssymb}
\usepackage{bbm}
\usepackage{latexsym}
\usepackage{lmodern}
\usepackage{mathrsfs}
\usepackage{slashed}
\usepackage{braket}
\usepackage{tensor}
\usepackage{simplewick}
\usepackage{graphicx}
\usepackage{xcolor}
\usepackage[font=small,labelfont=it]{caption}
\usepackage[font=scriptsize]{caption}
\usepackage[hidelinks]{hyperref}
\hypersetup{
        colorlinks=true,
        linkcolor=blue,
        citecolor=magenta,
        colorlinks = true,
        filecolor=magenta,      
        urlcolor=cyan,
        bookmarksnumbered=true,
        pdftitle={nicolai_sugra},
        bookmarks=true,
}
\usepackage[nottoc]{tocbibind} % Puts references in toc

\def\C{\mathbb C}

\newcommand{\al}{\alpha}
\newcommand{\be}{\beta}
\newcommand{\ga}{\gamma}
\newcommand{\de}{\delta}
\newcommand{\eps}{\epsilon}
\newcommand{\ve}{\varepsilon}

\newcommand{\ka}{\kappa}
\newcommand{\la}{\lambda}

\newcommand{\om}{\omega}

\newcommand{\sfrac}[2]{{\textstyle\frac{#1}{#2}}}

\newcommand{\pa}{\partial}
\newcommand{\im}{{\mathrm{i}}}

\newcommand{\ep}{{\mathrm{e}}}
\newcommand{\diff}{{\mathrm{d}}}

\newcommand{\beq}{\begin{equation}}
\newcommand{\eeq}{\end{equation}}
\newcommand{\eq}{\end{equation}}
\newcommand{\bea}{\begin{eqnarray}}
\newcommand{\eea}{\end{eqnarray}}
\newcommand{\with}{{\quad{\rm with}\quad}}

\renewcommand{\and}{{\quad{\rm and}\quad}}
\newcommand{\und}{{\qquad{\rm and}\qquad}}
\renewcommand{\=}{\ =\ }

\newcommand{\nn}{\nonumber}

\allowdisplaybreaks  % allow to break the content of align between page
\DeclareMathOperator{\tr}{tr}

\DeclareMathOperator{\Tr}{Tr}
\newcommand{\ie}{{i.e.}}
\newcommand{\eg}{{e.g.}}

\newcommand{\ltb}{\bigl\langle\!\!\!\bigl\langle\!\!\!\bigl\langle}
\newcommand{\rtb}{\bigr\rangle\!\!\!\bigr\rangle\!\!\!\bigr\rangle}

\newcommand{\Lagr}{ { \mathcal{L} } }
\newcommand{\Delt}{ { \mathcal{M} } }

\newcommand{\wLagr}{\widetilde{\Lagr}}
\newcommand{\wDelt}{\widetilde{\Delt}}
\newcommand{\wR}{\widetilde{R}}
\renewcommand{\wp}{\widetilde{\phi}}
\newcommand{\rR}{{\textrm{R}}}
\newcommand{\cov}{{\textrm{cov}}}

\advance \textheight by \topskip
\makeatletter
\@addtoreset{equation}{section}

\makeatother
\usepackage{tikz}
\usetikzlibrary{positioning,arrows}
\usetikzlibrary{decorations.pathmorphing}
\usetikzlibrary{decorations.markings}
\tikzset{
particle/.style={thick,draw=black, postaction={decorate},
    decoration={markings,mark=at position .50 with {\arrow[line width=1pt]{>}}  }},
aparticle/.style={thick,draw=black, postaction={decorate},
    decoration={markings,mark=at position .5 with  {\arrow[line width=1pt]{<}}  }},
gluon/.style={decorate, draw=black,
    decoration={coil,aspect=0}},
scalar/.style={thick,dashed,draw=black}
 }
\begin{document}

\begin{titlepage}
\setcounter{page}{0}

\phantom{.}
\vskip 1.5cm

\begin{center}

{\LARGE \bf 
Towards a Nicolai map for supergravity
}
\vspace{12mm}

{\Large Federico Arrighi, Saurish Khandelwal \ and \ Olaf Lechtenfeld}
\\[8mm]
\noindent {\em
Institut f\"ur Theoretische Physik and Riemann Center for Geometry and Physics\\
Leibniz Universit\"at Hannover, Appelstrasse 2, 30167 Hannover, Germany}
\vspace{12mm}

\begin{abstract} 
\noindent
We investigate the possibility of a Nicolai map for minimal supergravity in four dimensions.
Such a map would allow for the computation of quantum supergravity correlation functions in
terms of flat-space correlators in an effective nonlocal bosonic theory with the help of
a nonlinear field transformation, the inverse Nicolai map.  
Such a map is guaranteed for off-shell global supersymmetry, but local supersymmetry presents 
at least three obstacles for the construction. 
Their effects are analyzed in detail, in an attempt to set up a Nicolai map to leading order
in the gravitational coupling.
We find indications that the conformal factor of the metric obstructs the off-shell construction,
suggesting that the unimodular variant of supergravity may do better.
The on-shell supersymmetry approach, successful for super-Yang--Mills theory in its critical dimensions, 
also fails, because the graviton self-interaction cannot be written as a supervariation.
Nevertheless, by brute force we obtain a four-parameter first-order Nicolai map fulfilling 
the free-action condition.
For the acid test of determinant matching, however, one needs to push the general ansatz
and the perturbative expansion to the second order and to the quantum level.
\end{abstract}

\end{center}

\end{titlepage}
%

%\tableofcontents

\section{Introduction: the Nicolai map}\label{sec:intro}

\noindent
Every globally supersymmetric field theory features a (generically but not always)
nonlocal and nonlinear field transformation effecting a shift of its parameters,
say coupling constants~$g$. This so-called Nicolai map~$T$~\cite{Nic1,Nic2,Nic3}
relates the quantum expectation value of any functional~$Y$ built from the bosonic fields~$\phi$
at different $g$-values,
\begin{equation} \label{Tdef}
\bigl\langle \,Y[\phi]\, \bigr\rangle^\phi_g
\= \bigl\langle \,(T_{gg'}^{-1}Y)[\phi]\, \bigr\rangle^\phi_{g'}
\= \bigl\langle \,Y[T_{gg'}^{-1}\phi]\, \bigr\rangle^\phi_{g'}\ .
\end{equation}
Most importantly, this map allows one to compute such correlators in the free theory (at $g'{=}0$),
which is our reference coupling from now on.\footnote{
The vanishing vacuum energy implied by unbroken global supersymmetry normalizes $\langle1\rangle_g=1$.}
The value of the coupling is indicated by the subscript on the correlator and also on the symbol of 
the map~$T_g:\phi\mapsto \phi'[g,\phi]=T_g\phi$.
As made explicit in~\eqref{Tdef}, the Nicolai map is distributive, 
\ie~$T(\phi_1\phi_2)=(T\phi_1) \, (T\phi_2)$, 
It acts not in the original supersymmetric theory but in a bosonic nonlocal theory
defined by integrating out all other degrees of freedom, namely fermions~$\psi$
and possibly auxiliary fields~$F$, ghosts, Lagrange multipliers etc.,
\begin{equation}
\bigl\langle \,Y[\phi]\,\bigr\rangle^\phi_g \=
\bigl\langle\bigl\langle \,Y[\phi]\, \bigr\rangle\bigr\rangle^\phi_g \=
\ltb \,Y[\phi]\, \rtb_g\ ,
\end{equation}
where the inner bracket denotes the averaging over all fields except~$\phi$,
and the fat bracket applies to the original supersymmetric theory with all fields still present.
Hence, the expectation values of~\eqref{Tdef} denote a functional average over the remaining
bosonic fields in the effective nonlocal theory, governed by an action
\begin{equation} \label{Sloops}
S_g[\phi] \= S_g^{(0)}[\phi]\ +\ \sum_{r=1}^\infty \hbar^r\,S_g^{(r)}[\phi]\ ,
\end{equation}
where the classical local piece $S^{(0)}_g$ is the bosonic part of the original supersymmetric action
$S_{\textrm{\tiny SUSY}}[\phi,\psi,F]$ after eliminating auxiliaries, 
and the nonlocal quantum corrections~$S^{(r>0)}_g$ stem from the path integral over the fermions
in~$S_{\textrm{\tiny SUSY}}[\phi,\psi,F]$, all at coupling~$g$. 
The Feynman diagrammatic representation of~$S_g^{(r)}$ yields all graphs with $r$~fermion loops.

In case of fermion self-interactions in~$S_{\textrm{\tiny SUSY}}$ (as present in supergravity),
the Nicolai map is no longer classical but also a power series in~$\hbar$~\cite{CLR,L3},
\begin{equation} \label{quantmap}
T_g\phi \= T_g^{(0)}\!\phi \ +\ \sum_{r=1}^\infty \hbar^r\,T_g^{(r)}\!\phi\ ,
\end{equation}
in addition to the expansion in powers of~$g$. Also here,
$r$~denotes the number of fermion loops in a graphical expansion.
Substituting $Y\mapsto T_gY$ on the right-hand side of~\eqref{Tdef} and
comparing the path integrals, we derive the identity
\begin{equation} \label{matchingnew}
S^{(0)}_0[T_g\phi]\ +\ \sum_{r\ge1}\hbar^r\,S^{(r)}_0
\ -\ \im\hbar\,\Tr\ln\sfrac{\delta T_g\phi}{\delta\phi}
\= S^{(0)}_g[\phi]\ +\ \sum_{r\ge1} \hbar^r\,S_g^{(r)}[\phi]
\end{equation}
where $\Tr$ stands for the functional trace.
The terms $S_g^{(r>0)}$ lose their $\phi$~dependence at $g{=}0$
and hence the left-hand sum is a constant,
while $S^{(0)}_0$ is a quadratic functional of~$\phi$.
Inserting~\eqref{quantmap} into~\eqref{matchingnew} and separating powers of~$\hbar$
one arrives at an infinite hierarchy of `Nicolai-map conditions', one for each loop number. 
The tree-level and one-loop relations read
\begin{equation} \label{matching01}
S^{(0)}_0[T_g^{(0)}\!\phi] \= S^{(0)}_g[\phi] \quad\und\quad
S^{(0)}_0[T_g\phi]\big|_{O(\hbar)} + S^{(1)}_0 
- \im\,\Tr\ln\sfrac{\delta T_g^{(0)}\!\phi}{\delta\phi} \= S_g^{(1)}[\phi] \ .
\end{equation}
The first equation is known as the `free-action condition' and needs only the classical map,
while the second equation is the so-called `determinant-matching condition' modified by
a potentially nonzero~$T_g^{(1)}\phi$.\footnote{
Without this term, the left-hand side stems from the Jacobian determinant of the classical
Nicolai map, and the right-hand side comes from the fermion determinant
(due to the part quadratic in the fermions).}

There exists a construction method and a universal formula which produce a formal power series 
(in~$g$ and~$\hbar$) of the Nicolai map and its inverse~\cite{FL,DL1,L1,L2,ALMNPP,LR1,LR3,Lprague}.\footnote{
Sometimes the construction works also without off-shell supersymmetry,
\eg~for super Yang--Mills theory in dimensions 6 and~10 in the Landau gauge~\cite{ALMNPP,MN,LR2,LR4}.}
Its key ingredient is the so-called `coupling flow operator' 
\begin{equation}
R_g[\phi] \= \int\!\diff x\ \bigl(\pa_g T_g^{-1} \circ T_g \bigr) \phi(x)\,\frac{\delta}{\delta\phi(x)}\ ,
\end{equation}
where `$x$' stands for all coordinates our fields depend on.
The infinitesimal Nicolai map is governed by this functional differential operator,
\begin{equation} \label{localflow}
\pa_g\,\bigl\langle \,Y[\phi]\, \bigr\rangle^\phi_g 
\= \bigl\langle \,\bigl( \pa_g + R_g[\phi] \bigr) Y[\phi]\, \bigr\rangle^\phi_g\ ,
\end{equation}
also operating in the effective bosonic theory.
To obtain the coupling flow operator, 
one exploits off-shell global supersymmetry to write the original supersymmetric action as a supervariation,
\begin{equation} \label{defDelta}
S_{\textrm{\tiny SUSY}}[\phi,\psi,F] \= 
\int\!\diff x\ \delta_\alpha \Delt_\alpha[\phi,\psi,F](x)\ ,
\end{equation}
where $\alpha$ is a spinor index (we will be more concrete later) and the functional $\Delt_\alpha$ is 
the anticommuting penultimate component in the superfield expansion of the superspace action.
This starting point will require modification for {\em local\/} supersymmetry as we shall see.
Using~(\ref{defDelta}) and the supersymmetry Ward identity and integrating out all fields but~$\phi$
one finds the coupling flow operator as
\begin{equation} \label{flowop}
R_g[\phi] \= \frac{\im}{\hbar}\int\!\diff x\!\int\!\diff y\ 
\bigl\langle \,\pa_g\Delt_\alpha[\phi](y)\ 
\delta_\alpha \phi(x)\, \bigr\rangle\ \frac{\delta}{\delta\phi(x)}\ ,
\end{equation}
where the bracket without $\phi$ superscript indicates a functional averaging over all fermions,
auxiliary fields, ghosts, Lagrange multipliers etc.~in the supersymmetric theory.
For fields occurring only linearly under this bracket,
we are at this stage allowed to insert their on-shell values directly into this expression.
Sometimes only a fraction of the supersymmetry is needed for the construction,
which provides some flexibility in the sum over~$\al$ in~\eqref{flowop}.
This can then be employed to simplify the map.

In super Yang--Mills and supergravity theories, the supersymmetry transformations are nonlinear in the fields,
and therefore do not commute with $\pa_g$. One therefore needs to first scale out the coupling in front of the action
by absorbing it into the fields via a field rescaling~\cite{L1,LR1},
\begin{equation}
\widetilde{\phi} = g\,\phi\ ,\quad \widetilde{\psi} = g\,\psi\ , \quad\widetilde{F} = g\,F 
\qquad\Rightarrow\qquad
S_{\textrm{\tiny SUSY}}[\phi,\psi,F;g] \= \tfrac{1}{g^2}
\widetilde{S}_{\textrm{\tiny SUSY}}[\widetilde\phi,\widetilde\psi,\widetilde F;1]
\end{equation}
so that the coupling is fully explicit in
\begin{equation} \label{defDeltascaled}
\pa_g\,S_{\textrm{\tiny SUSY}}[\phi,\psi,F] \= -\frac{2}{g^3}
\int\!\diff x\ \delta_\alpha \wDelt_\alpha[\widetilde\phi,\widetilde\psi,\widetilde F](x)\ ,
\end{equation}
which produces a rescaled flow operator\footnote{
We do not add a tilde to the inner bracket or to $\delta_\alpha$; their meaning is obvious from the context.}
\begin{equation}
\wR[\widetilde\phi] \= -\frac{2}{g^2}\frac{\im}{\hbar}\int\!\diff x\!\int\!\diff y\ 
\bigl\langle \,\wDelt_\alpha[\widetilde\phi](y)\ 
\delta_\alpha \widetilde\phi(x) \bigr\rangle\ \frac{\delta}{\delta\widetilde\phi(x)}\ ,
\end{equation}
in a rescaled flow equation (with $\widetilde{Y}[\wp]=Y[\phi]$)
\begin{equation} \label{rescaledflow}
\pa_g\,\bigl\langle \,\widetilde{Y}[\wp]\,\bigr\rangle^{\wp}_g 
\= \bigl\langle \,\bigl( \pa_g + \tfrac1g\,\wR[\wp] \bigr)\,\widetilde{Y}[\wp]\,
\bigr\rangle^{\wp}_g\ .
\end{equation}
From this, the perturbative flow operator entering in~\eqref{localflow} is recovered by~\cite{LR1}
\begin{equation} \label{Euler}
\wR[\widetilde\phi{=}g\phi] \= E\ +\ g\,R_g[\phi] \quad\with\quad
E \equiv \int\!\diff x\ \phi(x)\,\frac{\delta}{\delta\phi(x)} \ ,
\end{equation}
where the degree-zero part $\wR[\widetilde\phi{=}g\phi]\big|_{g=0}$ must give the Euler operator~$E$
to allow for a perturbative expansion around~$g{=}0$.

The universal formula~\cite{LR1} directly represents the Nicolai map as the $g$-ordered exponential 
of ${-}\!\!\int_0^g\diff{g'}\,R_{g'}$. To extract a perturbative map, the exponential must be expanded,
hence the action of~$R_g$ has to be iterated, $R_{g_s\cdots}R_{g_2} R_{g_1}\phi$.
In case of a $k$-fermion self-interaction in~$S_\textrm{SUSY}$,
this iteration grafts full fermionic $k$-point functions onto previously produced diagrams.
For Wess--Zumino models and super-Yang--Mills theory ($k{=}2$) this generates fermionic trees only, 
dressed with bosonic `leaves', \ie~a classical Nicolai map.
For supersymmetric sigma models and supergravity ($k{=}4$), however, the graphical representation 
of the Nicolai map features a quartic fermion self-interaction and thus will involve fermionic
trees with all sorts of fermion loops embedded~\cite{CLR}. 

The present paper explores the possibility of a Nicolai map for minimal supergravity in four spacetime dimensions,
expanded around Minkowski spacetime. We encounter three obstacles in the off-shell construction of the map.
Due to being a density rather than a scalar the off-shell supergravity Lagrangian is only almost but not completely
expressible as a supervariation, as we show in Section~\ref{sec:action}. 
As a consequence in Section~\ref{sec:partialNicolai}, the flow equation picks up 
a multiplicative factor, and a potential Nicolai map will only be partial, \ie~accompanied by an additional 
measure factor in the path integral. In Section~\ref{sec:GF} we outline the gauge fixing and BRST quantization, which allows for 
the construction of a rescaled flow operator~$\wR$ and a second multiplicative correction in Section~\ref{sec:Rg}.
Armed with these expressions, Section~\ref{sec:RgPert} tests whether this flow operator at leading order in the gravitational coupling
reproduces the Euler operator, as is required (see~\eqref{Euler}) for an off-shell perturbative setup of the Nicolai map.
As the third obstruction we find that the test fails by a term proportional to the trace of the metric fluctuation. 
In Section~\ref{sec:onshell}, we abandon off-shell supersymmetry and try to build a Nicolai map following the super-Yang--Mills example
using only on-shell supersymmetry.
This approach necessarily generates multiplicative contributions in the flow equation~\eqref{localflow},
which may be taken into account as in Section~\ref{sec:partialNicolai}. Unfortunately, 
this candidate map does not fully produce the graviton self-interaction in the free-action condition to leading order.
However, relaxing its coefficients allows one to pass the free-action test with a four-parameter family,
providing a `brute-force' Nicolai map for supergravity to leading order in the gravitational coupling.
We also comment on the determinant-matching condition.  Finally, we conclude in Section~\ref{sec:outro}.
An Appendix \ref{sec:apx:onshell} presents a more general first-order ansatz with 21~terms 
and restricts them via the free-action condition.

\section{The action as a supervariation}\label{sec:action}

\noindent
Our goal is to construct an operator controlling the reaction of quantum supergravity correlators 
to a change of the gravitational coupling~$\ka$ in the minimal four-dimensional theory. 
To this end, we consider its off-shell locally supersymmetric action~\cite{PvN,FvP,DAZ}
\begin{equation} \label{sugraaction}
S_{\textrm{\tiny SUSY}} \= \int\!\diff^4x\ \Lagr_{\textrm{\tiny SUSY}} \= \int\!\diff^4x\ e\,\bigl\{
\tfrac{1}{2\ka^2}\,R\ -\ \tfrac12\,\bar\psi_\mu\ga^{\mu\nu\rho}D_\nu\psi_\rho\ -\ \tfrac13\,(S^2+P^2-A_\mu A^\mu)\bigr\}
\end{equation}
for the vierbein~$e^a_{\ \mu}$, its inverse~$e^\nu_{\ b}$ and the Majorana gravitino~$\psi_\mu^{\ \al}$ 
as well as an auxiliary axial vector~$A_\mu$, scalar~$S$ and pseudoscalar~$P$,
where we employ the standard abbreviations
\begin{align} &
e = \det(e^a_{\ \mu})\ ,\quad 
\ga^{\mu\nu\rho} = \ga^{[\mu}\ga^\nu\ga^{\rho]}\ ,\quad 
\ga^\mu = e^\mu_{\ a}\ga^a \ ,\quad
R = e^{a\mu} e^{b\nu} R_{\mu\nu a b}\bigl(\om(e,\psi)\bigr)\ , \\ &
D_\nu\psi_\rho = (\pa_\nu + \tfrac14\om_{\nu a b} \ga^{ab})\psi_\rho\ ,\quad
\om_{\nu a b} = \tfrac12 ( R_{ab,\nu} - R_{\nu a,b} + R_{\nu b,a} ), \quad
R_{\mu\nu,a} = -\pa_\mu e_{a\nu} + \pa_\nu e_{a\mu} +\tfrac12\ka^2\bar\psi_\mu\ga_a\psi_\nu\ ,\nn
\end{align}
freely converting indices with the (inverse) vierbein and lowering them with the spacetime metric
$g_{\mu\nu}=\eta_{ab}\,e^a_{\ \mu}e^b_{\ \nu}$ or the tangent (Minkowski) metric~$\eta_{ab}$.
Spinor indices~$\al,\be,\ldots$ are usually suppressed.
The gravitational coupling also appears in the deviation of the vierbein from the flat one,
\begin{equation}
e^a_{\ \mu} \= \de^a_{\ \mu}\ +\ \ka\,\phi^a_{\ \mu} \qquad\Rightarrow\qquad
g_{\mu\nu} \= \eta_{\mu\nu}\ +\ \ka(\phi_{\mu\nu}+\phi_{\nu\mu})\ +\ \ka^2\eta_{ab}\,\phi^a_{\ \mu}\phi^b_{\ \nu}\ .
\end{equation}
This action is off-shell invariant under diffeomorphisms, local Lorentz transformations and 
local supersymmetry transformations $\de_\eps=\eps_\al\de_\al$ with spinor parameter~$\eps$,\footnote{
For the construction we may restrict to the rigid part of supersymmetry, \ie~assume $\pa_\mu\eps=0$.}
\begin{equation}
\begin{aligned}
\tfrac1\ka\,\de_\eps e^a_{\ \mu} &\equiv \de_\eps \phi^a_{\ \mu} = \tfrac12\bar\eps\,\ga^a \psi_\mu \qquad\Rightarrow\qquad
\tfrac1\ka\,\de_\eps e^\mu_{\ a} = -\tfrac12\bar\eps\,\ga^\mu \psi_a \and \tfrac1\ka\,\de_\eps e = \tfrac12 e\,\bar\eps\,\ga^a\psi_a\ ,\\[4pt]
\de_\eps \psi_\mu &= \tfrac1\ka D_\mu\eps + \tfrac{\im}{2} A_\mu\ga_5\eps - \tfrac12\ga_\mu\,\eta\,\eps \quad\with
\eta = -\tfrac13( S - \im\ga_5 P - \im A_\rho\ga^\rho\ga_5)\ ,\\[4pt]
\de_\eps S &= \tfrac14 \bar\eps\,\ga^{\rho\sigma}\psi^\cov_{\rho\sigma}\ ,\quad
\de_\eps P = -\tfrac{\im}4 \bar\eps\,\ga_5 \ga^{\rho\sigma}\psi^\cov_{\rho\sigma}\ ,\quad 
\de_\eps A_\mu = \tfrac{\im}2\bar\eps\,\ga_5(\ga^\rho\psi^\cov_{\rho\mu}+\tfrac14\ga_\mu^{\ \rho\sigma}\psi^\cov_{\rho\sigma}) \\[2pt]
&\qquad\qquad\qquad\qquad\qquad\qquad\qquad\!\! \with
\tfrac12\psi^\cov_{\rho\sigma} = D_{[\rho}\psi_{\sigma]} + \tfrac{\im}2\ka A_{[\rho}\ga_5\psi_{\sigma]} - \tfrac12\ka\,\ga_{[\rho}\,\eta\,\psi_{\sigma]}\ ,
\end{aligned}
\end{equation}
implying
\begin{equation}
\tfrac{1}{\ka}\,\de_\eps \om_{\nu ab} \= \tfrac14\bar\eps\,(\ga_b\,\psi^\cov_{\nu a}-\ga_a\,\psi^\cov_{\nu b}-\ga_\nu\,\psi^\cov_{ab}
+\ka\ga_{ab}\,\eta\,\psi_\nu+\ka\,\eta\,\ga_{ab}\,\psi_\nu)\ .
\end{equation}

We want to obtain the off-shell lagrangian from the supersymmetry variation of
\begin{equation}
\tfrac{1}{4\ka}\,e\,\bar\ve\,\ga^{\mu\nu} \psi^\cov_{\mu\nu} \= \Delt_\ve^{\rm I} + \Delt_\ve^{\rm II} + \Delt_\ve^{\rm III}\ ,
\end{equation}
defining
\begin{equation}
\Delt_\ve^{\rm I} = \tfrac{1}{2\ka}\,e\,\bar\ve\,\ga^{\mu\nu}D_\mu\psi_\nu\ ,\qquad
\Delt_\ve^{\rm II} = \tfrac{\im}{4}\,e\,\bar\ve\,A^\mu\ga_5\psi_\mu\ ,\qquad
\Delt_\ve^{\rm III} = -\tfrac14\,e\,\bar\ve\,\ga^\mu(S-\im\ga^5 P)\psi_\mu\ \,
\end{equation}
where $\ve$ is another spinor parameter, and split it off via $\Delt_\ve=\bar\ve_\al\Delt_\al$.
A straightforward but lengthy computation yields
\begin{equation} \label{M123}
\begin{aligned}
\de_\al \Delt_\al^{\rm I} &\= 
\tfrac{1}{2\ka^2}\,e\,R -\tfrac{7}{16}\,e\,\bar\psi_\rho\ga^{\rho\mu\nu}D_\mu\psi_\nu -\tfrac14\,e\,\bar\psi_\mu\ga_\nu D^{[\mu}\psi^{\nu]}
+\tfrac{\im\ka}{32}\,e\,\bar\psi_\rho\ga^{\rho\mu\nu}A_\mu\ga_5\psi_\nu -\tfrac{\im\ka}{16}\,e\,\bar\psi_\mu\ga^{(\mu}A^{\nu)}\ga_5\psi_\nu \\
&\quad +\tfrac{\ka}{16}\,e\,\bar\psi_\mu(S-\im\ga_5 P)\psi^\mu\ ,\\[4pt]
\de_\al \Delt_\al^{\rm II} &\=
\tfrac13\,e\,A^2 -\tfrac{1}{16}\,e\,\bar\psi_\rho\ga^{\rho\mu\nu}D_\mu\psi_\nu +\tfrac14\,e\,\bar\psi_\mu\ga_\nu D^{[\mu}\psi^{\nu]}
-\tfrac{\im\ka}{32}\,e\,\bar\psi_\rho\ga^{\rho\mu\nu}A_\mu\ga_5\psi_\nu +\tfrac{\im\ka}{16}\,e\,\bar\psi_\mu\ga^{(\mu}A^{\nu)}\ga_5\psi_\nu \\
&\quad -\tfrac{\ka}{16}\,e\,\bar\psi_\mu(S-\im\ga_5 P)\psi^\mu\ ,\\[4pt]
\de_\al \Delt_\al^{\rm III} &\=
\tfrac23\,e\,(S^2+P^2) +\tfrac14\,e\,\bar\psi_\rho\ga^{\rho\mu\nu}D_\mu\psi_\nu +\tfrac12\,e\,\bar\psi_\mu\ga_\nu D^{[\mu}\psi^{\nu]}
+\tfrac{\im\ka}{8}\,e\,\bar\psi_\mu\ga^{(\mu}A^{\nu)}\ga_5\psi_\nu \\
&\quad -\tfrac{\ka}{8}\,e\,\bar\psi_\mu(S-\im\ga_5 P)\psi^\mu +\tfrac{\ka}{8}\,e\,\bar\psi_\mu\ga^{\mu\nu}(S-\im\ga_5 P)\psi_\nu\ .
\end{aligned}
\end{equation}
Clearly, $\Delt_\al^{\rm III}$ should be absent, so the best we can do is
\begin{equation} \label{best}
\begin{aligned}
&\Delt^{\rm inv} \= \Delt^{\rm I}+\Delt^{\rm II} \=
\tfrac{1}{2\ka}\,e\,\ga^{\mu\nu}D_\mu\psi_\nu \ +\ \tfrac{\im}{4}\,e\,A^\mu\ga_5\psi_\mu \\
\Rightarrow\quad
&\de_\al \Delt^{\rm inv}_\al \= e\,\bigl\{ \tfrac{1}{2\ka^2}\,R - \tfrac12\,\bar\psi_\mu\ga^{\mu\nu\rho}D_\nu\psi_\rho + \tfrac13\,A^2 \bigr\} 
\= \Lagr_{\textrm{\tiny SUSY}}\ +\ \tfrac13\,e\,(S^2+P^2)\ .
\end{aligned}
\end{equation}

This result is confirmed by a superspace calculation~\cite{wessbagger}. 
The failure to express the entire off-shell lagrangian as a supervariation
stems from the fact that it is the highest component not of a chiral superfield but of a chiral superfield {\sl density\/}
(and its hermitian conjugate). 
More concretely, in chiral superspace the action takes the form
\begin{align}
S_{\textrm{\tiny SUSY}} \= -6\int\!\diff^4x\,\diff^2\theta\ \mathcal{E}\,\mathcal{R} +\ \textrm{h.c.} 
\= -6 \int\!\diff^4x\ (\mathcal{E}\,\mathcal{R})|_{\theta^2} + \ \textrm{h.c.}
\end{align}
where $\mathcal{E}$ is the chiral density super-vielbein, and $\mathcal{R}$ is the chiral curvature superfield. 
The problem arises because the product $\mathcal{E}\,\mathcal{R}$ is itself a density superfield. 
When we look at its supersymmetry transformation, the chiral variation~\footnote{
In chiral superspace we work in Weyl spinor notation, with the supersymmetry transformation 
$\delta_\zeta = \zeta^{\alpha} \delta_{\alpha} + \bar{\zeta}_{\dot{\alpha}} \bar{\delta}^{\dot{\alpha}}$, 
where $\zeta^{\alpha}$ (chiral) and $\bar{\zeta}_{\dot{\alpha}}$ (anti-chiral) are the supersymmetry Weyl-spinor parameters.} 
of the $\theta$~component does not give only the desired term $(\mathcal{E}\,\mathcal{R})|_{\theta^2}$. 
Instead, it also brings in an extra contribution. Explicitly,
\begin{align}
\delta_{\zeta} \zeta  (\mathcal{E} \mathcal{R})|_{\theta} \= 
\zeta^\alpha \delta_{\alpha} \zeta^\beta  (\mathcal{E}\,\mathcal{R}|_{\theta})_\beta \= 
\zeta^2\bigl(-2 (\mathcal{E}\,\mathcal{R})|_{\theta^2} - \tfrac{2}{3} (M^*\,\mathcal{E}\,\mathcal{R})|_{\theta^0}\bigr)\ ,
\end{align}
where $M=S+\im P$.
Thus, the supervariation produces an extra contribution in agreement with our component result in~\eqref{best}. 
As a consequence of which, we shall encounter a multiplicative term in the flow equation.

\section{A partial Nicolai map}\label{sec:partialNicolai}

\noindent
Let us dwell a bit on the effect of a multiplicative term accompanying the coupling flow operator,
\begin{equation} 
\pa_\ka\,\bigl\langle \,Y[\phi]\, \bigr\rangle^\phi_\ka 
\= \bigl\langle \,\bigl( \pa_\ka + R_\ka[\phi] + Z_\ka[\phi] \bigr) Y[\phi]\, 
\bigr\rangle^\phi_\ka\ ,
\end{equation}
where $Z_\ka[\phi]\propto\tfrac{\im}{\hbar}\int\langle\!\langle e\,(S^2+P^2)\rangle\!\rangle$.
Integrating this relation gives a map relating gravitational correlators to flat-space ($\ka{=}0$) ones.
Suppressing the functional~$\phi$ dependence and inventing a similarity transformation, we obtain
\begin{equation}
\begin{aligned}
\bigl\langle \,Y\, \bigr\rangle^\phi_\ka &\=
\bigl\langle \,\exp\bigl\{ \ka\,(\pa_{\ka'}+R_{\ka'}+Z_{\ka'})\bigr\}\,Y\mid_{\ka'=0}\,
\bigr\rangle^\phi_0 \=
\bigl\langle \,\Sigma_{\ka'}^{-1}\,\exp\bigl\{ \ka\,(\pa_{\ka'}+R_{\ka'})\bigr\}\,\Sigma_{\ka'}\,Y\mid_{\ka'=0}\,
\bigr\rangle^\phi_0 \\[4pt] &\= 
\bigl\langle \,\bigl( \Sigma_0^{-1}\,\exp\bigl\{ \ka\,(\pa_{\ka'}+R_{\ka'})\bigr\}\,\Sigma_{\ka'}\bigr)_{\ka'=0}\
\bigl( \exp\bigl\{ \ka\,(\pa_{\ka'}+R_{\ka'})\bigr\}\,Y \bigr)_{\ka'=0}\, \bigr\rangle^\phi_0 \=
\bigl\langle \,U_\ka\ T_\ka^{-1}\,Y\, \bigr\rangle^\phi_0
\end{aligned}
\end{equation}
with an additional measure factor (abbreviating $\diff\equiv\pa_{\ka'}+R_{\ka'}$)
\begin{equation}
U_\ka \=  \Sigma_0^{-1}\,T_\ka^{-1}\,\Sigma_\ka \=
1\ +\ \ka\,Z_0\ +\ \tfrac1{2!}\ka^2 \bigl(Z_0^2+(\diff Z)_0\bigr)\ +\ \tfrac1{3!}\ka^3 \bigl(Z_0^3+3Z_0(\diff Z)_0+(\diff^2Z)_0\bigr)\ +\ \ldots
\end{equation}
collecting all terms containing the multiplicative piece~$Z$ or its derivatives.
The similarity factor is defined implicitly by
\begin{equation}
\pa_\ka+R_\ka+Z_\ka \= \Sigma_\ka^{-1}(\pa_\ka+R_\ka)\,\Sigma_\ka \qquad\Rightarrow\qquad
[\pa_\ka+R_\ka\,,\,\Sigma_\ka] \= \Sigma_\ka\,Z_\ka\ ,
\end{equation}
which was employed to derive the perturbative expansion of~$U_\ka$.
The extra measure factor~$U$ renders the vacuum energy $\ka$~dependent,
\begin{equation}
\ep^{\frac{\im}{\hbar}{\rm vol}\,{\cal E}_{\rm vac}} \= 
\ltb \,1\, \rtb_\ka \= 
\bigl\langle \,U_\ka\, \bigr\rangle^\phi_0\ ,
\end{equation}
which is actually familiar from supergravity, in contrast to super Yang--Mills theory.

\section{Gauge fixing}\label{sec:GF}

\noindent
As in any gauge theory, the computation of path integrals requires gauge fixing, which reduces the gauge to BRST invariance.
For supergravity, the local symmetry transformations comprise diffeomorphisms, local Lorentz transformations and local supersymmetry.
We choose a customary gauge fixing
\begin{equation}
0\= F_A \= \bigl( 
-e\,(\pa_\nu e^a_{\ \rho})(e^\rho_{\ a}\de^\nu_\mu-e_{a\mu}g^{\nu\rho}-e_a^{\ \nu}\de_\mu^\rho)\,,\,
\tfrac{\im}{2} (e_{ab}-e_{ba})\,,\,-\ga^\nu\psi_{\nu\al}\bigr)
\quad\with\ \ A = (\mu,ab,\al)\ ,
\end{equation}
combining all local indices into one. 
The diffeomorphisms are fixed by the harmonic or de Donder gauge $F_\mu=g_{\mu\nu}\pa_\rho(\sqrt{-g}\,g^{\nu\rho})$.
Expressing the Faddeev--Popov determinants in terms of ghost and antighost fields, we need to introduce the latter,
\begin{equation}
c^A \= \bigl( c^\mu\,,\,c^{ab}\,,\,c^\al\bigr) \und c^{*A} \= \bigl( c^{*\mu}\,,\,c^{*ab}\,,\,c^{*\al}\bigr)\ .
\end{equation}
We add Nakanishi--Lautrup auxiliary fields~$b^A$ to render the BRST transformations off-shell nilpotent.
The gauge-fixing and ghost part of the total lagrangian then reads
\begin{equation}
\Lagr_{\textrm{\tiny GF}} \= -\tfrac12 b^A\,T^{-1}_{AB}\,b^B + b^A F_A + c^{*A} 
\bigl(\tfrac{\pa F_A}{\pa e^a_{\ \mu}}\,s\,e^a_{\ \mu}+\tfrac{\pa F_A}{\pa\psi_\mu^{\ \al}}\,s\,\psi_\mu^{\ \al}\bigr)
\end{equation}
with our choice
\begin{equation}
T^{AB} \= \bigl( -\tfrac12 \de^{\mu\mu'}\,,\, \zeta\, \de^{aa'}\de^{bb'}\,,\, \xi\, \slashed\pa^{\al\al'} \bigr)
\end{equation}
of field-independent matrices ($\zeta$ and $\xi$ being quantum gauge parameters), and the Slavnov variations
\begin{equation}
\begin{aligned}
s\,e^a_{\ \mu} &\= \ka\,c^\nu\pa_\nu e^a_{\ \mu} + \ka\,e^a_{\ \nu}\pa_\mu c^\nu - \ka\,c^a_{\ b}e^b_{\ \mu}-\tfrac12\ka^2\bar\psi_\mu\ga^a c \ ,\\
s\,\psi_\mu^{\ \al} &\= \ka\,c^\nu\pa_\nu \psi_\mu^{\ \al} + \ka\,\psi_\nu^{\ \al}\pa_\mu c^\nu + 
(D_\mu c + \tfrac{\im\ka}{2}A_\mu\ga_5\,c - \tfrac{\ka}{2}\ga_\mu\eta\,c)^\al\ ,\\
s\,c^A &\= -\tfrac{\ka}{2}\,f^A_{\ BC}\,c^B c^C\ ,\qquad
s\,c^{*A} \= b^A\ ,\qquad
s\,b^A \= 0\ ,
\end{aligned}
\end{equation}
where $f^A_{\ BC}$ denote the structure constants of the gauge algebra, and the variations of the auxiliary fields~($A_\mu$, $S$, $P$) are not needed.

It is easy to see that $\Lagr_{\textrm{\tiny GF}}$ itself can be expressed as a Slavnov variation of a ``gauge-fixing fermion'',
\begin{equation}
\Lagr_{\textrm{\tiny GF}} \= s\,\Delt^{\rm gh} \qquad\with\qquad
\Delt^{\rm gh} \= 
c^{*A}\bigl( F_A - \tfrac12 T^{-1}_{AB} b^B \bigr)\ .
\end{equation}
It is important not to use prematurely the equation of motion for the Nakanishi--Lautrup auxiliary fields, $b^A=T^{AB}F_B$,
as this would render the procedure on-shell and prevent writing $\Lagr_{\textrm{\tiny GF}}$ as a Slavnov variation.

\section{Rescaled flow operator}\label{sec:Rg}

\noindent
In supersymmetric gauge theories, the coupling flow operator is not just a supervariation (see~\eqref{flowop}) but,
due to gauge fixing, needs to be improved by a Slavnov variation, which projects the flow onto the gauge slice.
It is found as follows,
\begin{equation}
\begin{aligned}
\pa_\ka\,\bigl\langle \,Y[\phi]\, \bigr\rangle^\phi_\ka \=
\pa_\ka\,\ltb\, Y[\phi] \,\rtb_\ka 
&\= \ltb\, \pa_\ka Y[\phi]\ +\ 
Y[\phi]\;\tfrac{\im}{\hbar}\pa_\ka(S_{\textrm{\tiny SUSY}}+S_{\textrm{\tiny GF}}) \,\rtb_\ka \\
&\= \ltb\, \pa_\ka Y[\phi]\ +\ 
Y[\phi]\;\tfrac{\im}{\hbar}\int\!\diff^4x\ \pa_\ka\bigl(\de_\al\Delt^{\rm inv}_\al+s\,\Delt^{\rm gh}-\tfrac{e}{3}(S^2{+}P^2)\bigr)\,\rtb_\ka \ ,
\end{aligned}
\end{equation}
Switching to rescaled fields (indices suppressed)~\cite{L1,LR1},\footnote{
In these references BRST transformations were used {\it on-shell\/} (without Nakanishi-Lautrup fields),
which necessitated a different ghost rescaling.}
\begin{equation} \label{rescale}
\widetilde{\phi} = \ka\,\phi\ ,\quad \widetilde{\psi} = \ka\,\psi\ , \quad
\widetilde{A} = \ka\,A\ , \quad \widetilde{S} = \ka\,S\ ,\quad \widetilde{P} = \ka\,P\ ,\quad
\widetilde{c} = \ka\,c\ ,\quad \widetilde{c}^* = \ka\,c^{*}\ ,\quad \widetilde{b} = \ka\,b\ ,
\end{equation}
to make the $\ka$~dependence explicit as a prefactor,
we can interchange \ $\pa_\ka\de_\al\tfrac{1}{\ka^2}\wDelt_\al=-\tfrac{2}{\ka^3}\de_\al\wDelt_\al$.
Now we are in a position to employ the supersymmetry and BRST Ward identities to find 
\begin{equation}
\begin{aligned}
\pa_\ka\,\ltb\, \widetilde{Y}[\wp] \,\rtb_\ka
&\= \ltb\, \pa_\ka \widetilde{Y}[\wp] 
\ -\ \tfrac{2}{\ka^3}\,\tfrac{\im}{\hbar}\int\!\diff^4x\ 
\bigl(\wDelt^{\rm inv}_\al\de_\al+\wDelt^{\rm gh}\,s-\tfrac{e}{3}(\widetilde{S}^2{+}\widetilde{P}^2)\bigr)\,\widetilde{Y}[\wp] \\
&\qquad\qquad\qquad -\ \tfrac{2}{\ka^5}\,\tfrac{\im}{\hbar}\int\!\diff^4x\ 
\wDelt^{\rm inv}_\al\ \tfrac{\im}{\hbar}\int\!\diff^4x'\ \bigl(\de_\al\,s\,\wDelt^{\rm gh}\bigr)\ \widetilde{Y}[\wp]\, \rtb_\ka \\
&\= \ltb\, \pa_\ka \widetilde{Y}[\wp]
\ -\ \tfrac{2}{\ka^3}\,\tfrac{\im}{\hbar}\int\!\diff^4x\
\bigl(\wDelt^{\rm inv}_\al\de_\al+\wDelt^{\rm gh}\,s\bigr)\,\widetilde{Y}[\wp] \\
&\qquad\qquad\qquad -\ \tfrac{2}{\ka^5}\,\tfrac{\im}{\hbar}\int\!\diff^4x\ 
\wDelt^{\rm inv}_\al\,\tfrac{\im}{\hbar}\int\!\diff^4x'\ \bigl(\de_\al\wDelt^{\rm gh}\bigr)\,s\,\widetilde{Y}[\wp] \\
&\qquad\qquad\qquad -\ \tfrac{2}{\ka^5}\,\tfrac{\im}{\hbar}\int\!\diff^4x\
\wDelt^{\rm inv}_\al\ \tfrac{\im}{\hbar}\int\!\diff^4x'\ \bigl(\{\de_\al,s\}\,\wDelt^{\rm gh}\bigr)\ \widetilde{Y}[\wp] \\
&\qquad\qquad\qquad +\ \tfrac{2}{\ka^3}\,\tfrac{\im}{\hbar}\int\!\diff^4x\ 
\tfrac{e}{3}(\widetilde{S}^2{+}\widetilde{P}^2)\bigr)\,\widetilde{Y}[\wp] \, \rtb_\ka \\
&\ :=\ \bigl\langle\,\pa_\ka \widetilde{Y}[\wp] 
\ +\ \tfrac1\ka\,\bigl(\wR^{\rm inv}+\wR^{\rm gh}\bigr)\,\widetilde{Y}[\wp]
\ +\ \tfrac1{\ka}\,\wR^{\rm mix}\,\widetilde{Y}[\wp] 
\ +\ \tfrac1{\ka}\,\widetilde{Z}[\wp]\ \widetilde{Y}[\wp]\, \bigr\rangle^\phi_\ka\ ,
\end{aligned}
\end{equation}
where we used that
\begin{equation}
\de_\al \wLagr_{\textrm{\tiny SUSY}} =0 \und
s\, \wLagr_{\textrm{\tiny GF}} =0 \ ,
\end{equation}
and we defined 
\begin{equation}
\begin{aligned}
\wR^{\rm inv} &\= -\tfrac{2}{\ka^2}\,\tfrac{\im}{\hbar}\int\!\diff^4x\ \bigl\langle\wDelt^{\rm inv}_\al\,\de_\al\bigr\rangle
\= -\tfrac{2}{\ka^2}\,\tfrac{\im}{\hbar}\int\!\diff^4x\ \bigl\langle\wDelt^{\rm inv}_\al\,
\int\!\diff^4x'\ (\de_\al e^a_{\ \mu})\bigr\rangle\,\tfrac{\de}{\de e^a_{\ \mu}}\ ,\\
\wR^{\rm gh} &\= -\tfrac{2}{\ka^2}\,\tfrac{\im}{\hbar}\int\!\diff^4x\ \bigl\langle\wDelt^{\rm gh}\,s\bigr\rangle
\= -\tfrac{2}{\ka^2}\,\tfrac{\im}{\hbar}\int\!\diff^4x\ \bigl\langle 
\tfrac12\widetilde{F}_A\,\widetilde{c}^{*A}\,\int\!\diff^4x'\,(s\,e^a_{\ \mu})\bigr\rangle\,\tfrac{\de}{\de e^a_{\ \mu}}\ ,\\
\wR^{\rm mix} & \= \tfrac{2}{\ka^4}\,\tfrac{1}{\hbar^2}\int\!\diff^4x\ \bigl\langle
\wDelt^{\rm inv}_\al\,\int\!\diff^4x'\ (\de_\al\widetilde{F}_A)\,\widetilde{c}^{*A}\,
\int\!\diff^4x''\,(s\,e^a_{\ \mu})\bigr\rangle\,\tfrac{\de}{\de e^a_{\ \mu}}\ , \\
\widetilde{Z} &\= -\ \tfrac{2}{\ka^4}\,\tfrac{\im}{\hbar}\int\!\diff^4x\
\bigl\langle\wDelt^{\rm inv}_\al\ \tfrac{\im}{\hbar}\int\!\diff^4x'\,\widetilde{c}^{*A}\{\de_\al,s\}\,\widetilde{F}_A\bigr\rangle
\ +\ \tfrac{2}{\ka^2}\,\tfrac{\im}{\hbar}\int\!\diff^4x\ \tfrac{e}{3}\bigl\langle\widetilde{S}^2{+}\widetilde{P}^2\bigr\rangle\ .
\end{aligned}
\end{equation}
The $\wR^{\rm mix}\widetilde{Y}$ term assures that the flow remains on the bosonic gauge slices, because 
\begin{equation}
\begin{aligned}
\smallint\bigl\langle \widetilde{F}_A\,\widetilde{c}^{*A}\,s\,\widetilde{F}_B\bigr\rangle = 
\im\ka^2 \widetilde{F}_B  \qquad & \Rightarrow\qquad
\wR^{\rm gh}\widetilde{F}_B = \widetilde{F}_B  \quad\und\quad
(\wR^{\rm inv} + \wR^{\rm mix})\,\widetilde{F}_B =0 \\ & \Rightarrow\qquad
\bigl(\pa_\ka+\tfrac1\ka(\wR^{\rm inv}+\wR^{\rm gh}+\wR^{\rm mix})\bigr)\,\tfrac1\ka\widetilde{F}_B =0
\end{aligned}
\end{equation}
for $B\in\{\mu,ab\}$,
and the multiplicative factor $\widetilde{Z}$ is harmless in this respect.
In globally supersymmetric gauge theories, $\{\de_\al,s\}=0$, and $\widetilde{Z}=0$, so one obtains a proper (differential) flow operator.
Supergravity, however, seems to require a multiplicative factor in the flow equation, allowing for a partial Nicolai map only.

\section{A perturbative flow operator?}\label{sec:RgPert}

\noindent
In order to set up the perturbation expansion for a (partial) Nicolai map, we have to undo the rescaling~\eqref{rescale}
and expand~\footnote{
Note the difference between $\rR_i$ and $R_\ka$ as well as between $\textrm{Z}_i$ and $Z_\ka$. In particular, $\rR_1=R_0$.}
\begin{equation} 
\begin{aligned}
\wR[\wp{=}\ka\phi] &\= \rR_0[\phi] + \ka\,\rR_1[\phi] + \ka^2 \rR_2[\phi] + \ldots 
\ \buildrel{!}\over{=}\ E\ +\ \ka\,R_\ka[\phi] \ ,\\
\widetilde{Z}[\wp{=}\ka\phi] &\= \textrm{Z}_0[\phi] + \ka\,\textrm{Z}_1[\phi] + \ka^2 \textrm{Z}_2[\phi] + \ldots 
\ \buildrel{!}\over{=}\ 0\ +\ \ka\,Z_\ka[\phi]\ ,
\end{aligned}
\end{equation}
where the leading term is fixed by regularity at $\ka{=}0$.
Let us first observe that a seemingly more singular term in $\wR^{\rm inv}$ and $\widetilde{Z}$ is absent,
\begin{equation}
-\tfrac{2}{\ka}\,\tfrac{\im}{\hbar}\int\!\diff^4x\ \tfrac12\,e\,\ga^{\mu\nu}D_\mu\psi_\nu
\= -\tfrac{2}{\ka}\,\tfrac{\im}{\hbar}\int\!\diff^4x\ \tfrac12\,\bigl\{\pa_a(\ga^{ab}\psi_b) + O(\ka)\bigr\} \= O(\ka^0)\ ,
\end{equation}
because we may drop a total derivative.
Therefore, $\wDelt^{\rm inv}_\al$, $\wDelt^{\rm gh}$ and $\widetilde{Z}$ begin with order $\ka^2$.
Indeed, $\textrm{Z}_0=0$, because not only provide $\widetilde{c}^{*A}$, $\widetilde{F}_A$, $\widetilde{S}$ and $\widetilde{P}$
a factor of~$\ka$ each upon scaling back, but the graded commutator~$\{\de_\al,s\}$ produces a structure constant of the gauge
algebra, which carries~$\ka$, and the auxiliary-field equations of motion
\begin{equation}
e\,S = \ka\,\bar{c}\,c\ ,\qquad
e\,P = \im\ka\,\bar{c}\,\ga_5 c\ ,\qquad
e\,A_\mu = \tfrac{\im}{4}\ka\,\bar{c}\,\ga_5\ga_\mu c
\end{equation}
yield further factors of~$\ka$.

Next, we check for $R_0[\phi]=E$ from
\begin{equation}
\begin{aligned}
\wR^{\rm inv} &\=  -\tfrac{2\im}{\hbar}\int\!\diff^4x\ \bigl\langle\bigl\{
\tfrac{1}{2\ka}\,e\,\ga^{\mu\nu}(\pa_\mu+\tfrac14\omega_{\mu ab}\ga^{ab})\psi_\nu + \tfrac{\im}{4}\,e\,A^\mu\ga_5\psi_\mu \bigr\}_\al
\int\!\diff^4x'\ \bigl(-\tfrac12\bar\psi_\rho\ga^e\bigr)_\al\bigr\rangle\,\tfrac{\de}{\de\phi^e_{\ \rho}}\ ,\\
\wR^{\rm gh} &\= -\tfrac{2\im}{\hbar}\int\!\diff^4x\ \bigl\langle\tfrac12\,c^{*A}\,F_A
\int\!\diff^4x'\ \bigl(c^\nu\pa_\nu e^e_{\ \rho} + e^e_{\ \nu}\pa_\rho c^\nu - c^{ed}e_{d\rho}-\tfrac12\ka\,\bar\psi_\rho\ga^e c\bigr)\bigr\rangle\,
\tfrac{\de}{\de\phi^e_{\ \rho}} \ ,\\
\wR^{\rm mix} &\= - \wR^{\rm inv}\,\wR^{\rm gh} \ +\ O(\ka)\ .
\end{aligned}
\end{equation}
To this end, we expand (abbreviating $\phi=\phi^a_{\ a}$ and writing $\approx$ when discarding $O(\ka^2)$ terms)
\begin{equation}
\begin{aligned}
e_{a\mu} &\= \de_{a\mu}+\ka\,\phi_{a\mu} \qquad\Rightarrow\qquad
e_{\mu a} \ \approx\ \de_{\mu a}+\ka\,\phi_{\mu a} \ ,\quad
e^\mu_{\ a}\ \approx\ \de^\mu_a-\ka\,\phi^\mu_{\ a}\ ,\quad
e \ \approx\ 1 + \ka\,\phi\ ,\\
\omega_{\mu ab} &\ \approx\ \tfrac12\ka\,\bigl( \pa_b(\phi_{a\mu}+\phi_{\mu a}) - \pa_a(\phi_{b\mu}+\phi_{\mu b}) - \pa_\mu(\phi_{ab}-\phi_{ba})\bigr)
\end{aligned}
\end{equation}
and (after applying all variations) impose the gauge conditions
\begin{equation} \label{gaugephi}
\pa^\mu(\phi_{a\mu}{+}\phi_{\mu a})=\pa_a\phi \und
\phi_{a\mu}=\phi_{\mu a}\ ,
\end{equation}
which simplifies
\begin{equation}
\tfrac1{4\ka}\,\omega_{\mu ab}\,\ga^{\mu\nu}\ga^{ab}\ \approx\
-\tfrac14\pa^\nu\phi+\tfrac14\pa_c\phi\,\ga^{c\nu}-\tfrac12\pa_c\phi^\nu_{\ d}\ga^{cd}\ .
\end{equation}
Using the gauge-fixed free gravitino propagator~\cite{baulieu}
\begin{equation} \label{psiprop}
\tfrac1{\im\hbar}\,\bigl\langle \psi_{\nu\al}\ \bar\psi_{\rho\beta} \bigr\rangle_{\ka=0} \= 
\bigl(\tfrac12\,\ga_\rho\ga_\la\ga_\nu+\tfrac{1-2\xi}{\xi}\,\pa_\rho\ga_\la\pa_\nu/\Box\bigr)_{\al\be} \tfrac{\pa^\la}{\Box}
\end{equation}
we obtain
\begin{align} \label{R0inv}
\rR_0^{\rm inv}&\= \tfrac{\im}{\hbar}\int\!\diff^4x\ \bigl\langle\bigl\{ 
\tfrac1{2\ka}(1+\ka\,\phi)(\de^\mu_c-\ka\,\phi^\mu_{\ c})(\de^\nu_d-\ka\,\phi^\nu_{\ d})\ga^{cd}
(\pa_\mu+\tfrac14\omega_{\mu ab}\ga^{ab})\,\psi_\nu\bigr\}_\al
\int\!\diff^4x'\ \bigl(\bar\psi_\rho\ga^e\bigr)_\al\bigr\rangle_{\ka=0}\,\tfrac{\de}{\de\phi^e_{\ \rho}} \nn\\
&\= -\tfrac{1}{2}\int\!\!\!\int\!\diff^4x\,\diff^4x' \tr\,\bigl\{ 
\bigl( \phi\,\ga^{\mu\nu}+2\phi^{[\mu}_{\ \ c}\ga^{\nu]c}\bigr)\,\pa_\mu
-\tfrac14\bigl(\pa^\nu\phi-\pa_c\phi\,\ga^{c\nu}+2\pa_c\phi^\nu_{\ d}\ga^{cd}\bigr)\bigr\} \nn\\
& \qquad\qquad\qquad\qquad\qquad\qquad\qquad\qquad\qquad\qquad
\times\,\bigl\{ \tfrac12\,\ga_\rho\ga_\la\ga_\mu +\tfrac{1-2\xi}{\xi}\,\pa_\rho\ga_\la\pa_\nu/\Box\bigr\}\,
\ga^e\,\tfrac{\pa^\la}{\Box}\,\tfrac{\de}{\de\phi^e_{\ \rho}} \\
&\=  -\tfrac{1}{2}\int\!\!\!\int\!\diff^4x\,\diff^4x' \tr\,\bigl\{ 
\tfrac14\phi\,\eta^{\mu\nu}+\tfrac34\phi\,\ga^{\mu\nu}+2\phi^{[\mu}_{\ \ c}\ga^{\nu]c}+\tfrac12\phi^\nu_{\ d}\ga^{\mu d}\bigr\} \nn\\
& \qquad\qquad\qquad\qquad\qquad\qquad\qquad\qquad\qquad\qquad
\times\,\bigl\{ \tfrac12\,\ga_\rho\ga_\la\ga_\mu +\tfrac{1-2\xi}{\xi}\,\pa_\rho\ga_\la\pa_\nu/\Box\bigr\}\,
\ga^e\,\tfrac{\pa_\mu\pa^\la}{\Box}\,\tfrac{\de}{\de\phi^e_{\ \rho}} \nn\\
&\=  \int\!\!\!\int\!\diff^4x\,\diff^4x' \bigl\{ 
\phi^e_{\ \rho} - \tfrac14\phi\,\de^e_\rho - \tfrac{1}{2\xi}\,\phi\,\tfrac{\pa^e\pa_\rho}{\Box} \bigr\}\,\tfrac{\de}{\de\phi^e_{\ \rho}} \nn
\end{align}
by partially integrating and performing the gamma traces indicated by `tr'.

Regarding the ghost contribution, we need the free diffeomorphism and Lorentz ghost propagators 
\begin{equation}
\tfrac1\hbar\,\bigl\langle c^{*\mu}\ c^\nu \bigr\rangle_{\ka=0} \= -\im\,\tfrac{\eta^{\mu\nu}}{\Box} \ ,\qquad
\tfrac1\hbar\,\bigl\langle c^{*ab}\ c^{cd} \bigr\rangle_{\ka=0} \= 
\tfrac12 (\eta^{ad}\eta^{bc}-\eta^{ac}\eta^{bd}) \ ,\qquad
\tfrac1\hbar\,\bigl\langle c^{*\mu}\ c^{cd} \bigr\rangle_{\ka=0} \= -\im\,\tfrac{\eta^{\mu[c}\pa^{d]}}{\Box}
\end{equation}
and find
\begin{align}
\rR_0^{\rm gh}&\= -\tfrac{\im}{\hbar}\int\!\diff^4x\ \bigl\langle\bigl\{ 
c^{*\mu} (-\pa_\mu\phi+\pa_a\phi^a_{\ \mu}+\pa_a\phi_\mu^{\ a})+ \tfrac{\im}{2} c^{*ab}(\phi_{ab}-\phi_{ba})\bigr\}
\int\!\diff^4x'\ \bigl( \de^e_\nu\,\pa_\rho c^\nu - \eta_{d\rho}\,c^{ed}\bigr)\bigr\rangle_{\ka=0}\,\tfrac{\de}{\de\phi^e_{\ \rho}} \nn\\
&\= \int\!\!\!\int\!\diff^4x\,\diff^4x' \bigl\{ 
(\pa_\mu\phi-\pa_a\phi^a_{\ \mu}-\pa_a\phi_\mu^{\ a})\,\tfrac12(\eta^{\mu e}\pa_\rho+\de^\mu_\rho\pa^e)\,\tfrac1\Box 
-\tfrac{1}{4}\, (\phi_{ab}-\phi_{ba})(\de^a_\rho\eta^{be}-\eta^{ae}\de^b_\rho)\bigr\}\,
\tfrac{\de}{\de\phi^e_{\ \rho}} \nn\\
&\= \int\!\!\!\int\!\diff^4x\,\diff^4x' \bigl\{ -\phi\,\tfrac{\pa^e\pa_\rho}{\Box}
+\tfrac12(\phi^{ae}+\phi^{ea})\,\tfrac{\pa_a\pa_\rho}{\Box}+\tfrac12(\phi^a_{\ \rho}+\phi_\rho^{\ a})\,\tfrac{\pa_a\pa^e}{\Box}\,
\;+\; \tfrac12\,(\phi^e_{\ \rho}-\phi_\rho^{\ e}) \bigr\}\,\tfrac{\de}{\de\phi^e_{\ \rho}} \ ,
\end{align}
which indeed obeys
\begin{equation}
\rR_0^{\rm gh} F_\mu = F_\mu \quad\und\quad \rR_0^{\rm gh} F_{ab} = F_{ab}\ .
\end{equation}
After partial integrations, this operator vanishes on the gauge slice defined by~\eqref{gaugephi}, as it should.

Finally, the `mixed' part, evaluated on the gauge slice, turns out to be
\begin{align}
& \rR_0^{\rm mix} \nn \\
&\=  -\int\!\!\!\int\!\!\!\int\!\!\!\int\!\diff^4x\,\diff^4x'\,\diff^4y\,\diff^4y' \bigl\{ 
\phi^d_{\ \nu} - \tfrac14\phi\,\de^d_\nu - \tfrac{1}{2\xi}\,\phi\,\tfrac{\pa^d\pa_\nu}{\Box} \bigr\}\,\tfrac{\de}{\de\phi^d_{\ \nu}} 
\bigl\{-\phi\,\tfrac{\pa^e\pa_\rho}{\Box} + \phi^{ae}\,\tfrac{\pa_a\pa_\rho}{\Box} + \phi_{a\rho}\,\tfrac{\pa^a\pa^e}{\Box}\bigr\}
\,\tfrac{\de}{\de\phi^e_{\ \rho}} \nn\\
&\=  -\int\!\!\!\int\!\!\!\int\!\diff^4x\,\diff^4x'\,\diff^4y \bigl\{ 
\phi^d_{\ \nu} - \tfrac14\phi\,\de^d_\nu - \tfrac{1}{2\xi}\,\phi\,\tfrac{\pa^d\pa_\nu}{\Box} \bigr\}\,
\bigl\{-\de^\nu_d\,\tfrac{\pa^e\pa_\rho}{\Box} + \eta^{\nu e}\,\tfrac{\pa_d\pa_\rho}{\Box} + \de^\nu_\rho\,\tfrac{\pa_d\pa^e}{\Box}\bigr\}
\,\tfrac{\de}{\de\phi^e_{\ \rho}} \\
&\=  -\int\!\!\!\int\!\diff^4x\,\diff^4x' \bigl\{
-\tfrac{1+\xi}{2\xi}\,\phi\,\pa^e\pa_\rho + \phi^{de}\,\pa_d\pa_\rho + \phi^d_{\ \rho}\,\pa_d\pa^e \bigr\}\,
\tfrac1\Box\,\tfrac{\de}{\de\phi^e_{\ \rho}} \nn\\
&\= \int\!\!\!\int\!\diff^4x\,\diff^4x'\ \tfrac{1-\xi}{2\xi}\phi\,\tfrac{\pa^e\pa_\rho}{\Box}\,\tfrac{\de}{\de\phi^e_{\ \rho}} \nn\ ,
\end{align}
where in the last step we partially integrated and used~\eqref{gaugephi}.
We learn that~$\rR_0^{\rm mix}$ removes the gauge-dependent part of~$\rR_0^{\rm inv}$ in~\eqref{R0inv}. 
In total, on the gauge slice we arrive at
\begin{equation}
\rR_0 \= \int\!\diff^4x\ \bigl\{ \phi^e_\rho\,\tfrac{\de}{\de\phi^e_{\ \rho}} 
\ -\ \tfrac14\,\phi\,\tfrac{\de}{\de\phi} \ -\ \tfrac12\,\phi\,\tfrac{\pa^e\pa_\rho}{\Box}\,\tfrac{\de}{\de\phi^e_{\ \rho}} \bigr\}\ ,
\end{equation}
of which only the first term provides the required Euler operator!
We are forced to conclude that the off-shell perturbative construction of the Nicolai map fails for minimal supergravity.

\section{On-shell approach}\label{sec:onshell}

\noindent
In super-Yang--Mills theory in the Landau gauge, it is possible to construct the Nicolai map using only on-shell supersymmetry
and avoiding the rescaling. Since $\pa_g\Lagr_{\textrm{\tiny SUSY}}$ can only incompletely be expressed as a supervariation,
also there we encounter multiplicative $Z$~terms in the flow operator which, however, ultimately cancel in the critical 
spacetime dimensions. This is not to be expected for supergravity. Still, let us investigate whether this example
carries over to a (potentially partial) Nicolai map for supergravity, at least to leading order in~$\ka$.

We recapitulate the situation in the super-Yang--Mills case~\cite{ALMNPP}, 
for a Lie-algebra valued Yang--Mills potential~$A_\mu$ and Majorana gaugino~$\lambda\in\C^r$ 
in the Landau gauge $\pa{\cdot}A=0$ with a gauge coupling~$g$. 
The $g$-derivative of the on-shell supersymmetric lagrangian is not a superfield component, but still we may write
\begin{equation} \label{SYMonshell}
\begin{aligned}
\pa_g \tr\,\bigl\{ -\tfrac14\,F_{\mu\nu}F^{\mu\nu} -\tfrac{\im}{2}\,\bar\lambda\,\slashed{D}\lambda \bigr\} &\=
2\,\delta_\al\pa_g\Delt_\al\ +\ \bigl(\tfrac{D-1}{r}-\tfrac12\bigr)\,\tr\,\bigl\{\im\,\bar\lambda\,\slashed{A}{\times}\lambda\bigr\} \\[2pt]
\textrm{with}\qquad
\Delt_\al \=  -\tfrac1{4r}\,\tr\,\bigl\{ F_{\mu\nu}\,(\ga^{\mu\nu}\lambda)_\al\bigr\}
\qquad &\Rightarrow \qquad 
\pa_g\Delt_\al \= -\tfrac1{4r}\,\tr\,\bigl\{ A_\mu{\times}A_\nu\,(\ga^{\mu\nu}\lambda)_\al\bigr\}\ ,
\end{aligned}
\end{equation}
where `${\times}$' indicates a contraction with the Lie-algebra structure constants and `tr' refers to the color trace.
We notice that, even though $\pa_g$ does not commute with $\de_\al$, acting in different order on $\Delt_\al$ gives
the same bosonic interaction $-\tfrac12\tr\{A_\mu{\times}\!A_\nu\,F^{\mu\nu}\}$ up to a factor of~$2$.\footnote{
The auxiliary field~$D$ does not contribute to this argument.}
The last term in the upper line of~\eqref{SYMonshell} depends on the spacetime dimension~$D$ and the Majorana spinor dimension~$r$.
It contributes to a multiplicative factor $Z_g$ in the flow equation, as does $\pa_g{\cal L}_{\textrm{\tiny GF}}$
and another term generated from the supersymmetric Ward identity. The total $Z_g$ factor turns out to cancel if and only if
\begin{equation}
\tfrac{D-1}{r}-\tfrac12=\tfrac1r \qquad\Leftrightarrow\qquad r=2(D{-}2) \qquad\Leftrightarrow\qquad D=3,4,6,10\ ,
\end{equation}
which are precisely the critical dimensions in which pure super-Yang--Mills theory is known to exist.
The leading order (in~$g$) of the flow operator is not affected by the multiplicative modification and takes the form
\begin{equation}
\begin{aligned}
R_g^{\rm inv} &\= \tfrac{2\im}{\hbar} \int\!\diff^4y\ \bigl\langle\,\pa_g\Delt_\al(y)\ \de_\al\,\bigr\rangle 
\= -\tfrac{2\im}{\hbar} \int\!\diff^4y\int\!\diff^4x\ 
\bigl\langle\,\pa_g\Delt_\al(y)\ (\bar\lambda\,\ga_\mu)_\al(x)\,\bigr\rangle\,\tfrac{\de}{\de A_\mu(x)}\\
&\= \tfrac{\im}{2r\hbar}\tr\int\!\diff^4y\int\!\diff^4x\
(A_\rho{\times}A_\sigma)(y)\,\gamma^{\rho\sigma} \bigl\langle\,\lambda(y)\ \bar\lambda(x)\,\bigr\rangle\,\gamma_\mu\,\tfrac{\de}{\de A_\mu(x)} \\
&\= -\int\!\diff^4y\int\!\diff^4x\ (A_\mu{\times}A_\nu)(y)\,\tfrac{\pa^\nu}{\Box}(y{-}x)\,\tfrac{\de}{\de A_\mu(x)}\ +\ O(g)\ .
\end{aligned}
\end{equation}
with $\tr 1=r$ and $\tfrac1\hbar\langle\lambda(y)\,\bar\lambda(x)\rangle_{g=0}=\im\frac{\slashed\pa}{\Box}$.
Therefore, the Nicolai map starts out as~\cite{Nic2}
\begin{equation}
T_g A_\mu(x) \= A_\mu(x)\ -\ g\,R_{g=0}^{\rm inv}A_\mu(x)\ +\ O(g^2) \=
A_\mu(x)\ -\ g\int\!\diff^4y\ \tfrac{\pa^\nu}{\Box}(x{-}y)\,(A_\mu{\times}A_\nu)(y) \ +\ O(g^2)\ .
\end{equation}
Sticking this into the free action, one obtains
\begin{equation}
\begin{aligned}
\tfrac12\int\!\diff^4x\ T_g A^\mu(x)\,\Box\,T_g A_\mu(x) 
&\=\tfrac12\int\!\diff^4x\ A^\mu(x)\,\Box\,A_\mu(x) \ -\ g\int\!\diff^4x\ A^\mu\,\pa^\nu(A_\mu{\times}A_\nu)(x)\ +\ O(g^2)\\
&\=\tfrac12\int\!\diff^4x\ A^\mu(x)\,\Box\,A_\mu(x) \ -\ g\int\!\diff^4x\ (\pa^\mu A^\nu)\,(A_\mu{\times}A_\nu)(x)\ +\ O(g^2)\ ,
\end{aligned}
\end{equation}
which produces the correct cubic part of the Yang--Mills lagrangian.

In supergravity, we impose the de Donder gauge for the diffeomorphisms and remove the asymmetric part of the vierbein.
Expanding around Minkowski space to leading order in~$\ka$,
we may convert world into tangent indices and thus work with a field
\begin{equation} \label{landau}
\phi_{ab} \qquad\textrm{subject to}\qquad
\pa^a\phi_{ab} = \tfrac12\pa_b\phi \und \phi_{ab}=\phi_{ba} \qquad\textrm{with}\quad \phi\equiv\phi_{ab}\eta^{ab}\ ,
\end{equation}
hence $\phi_{ab}$ is symmetric and need not be Lorentz-contracted with a derivative.
We are seeking the leading order (in~$\ka$) of the classical part of a Nicolai map,
\begin{equation} \label{mapleading}
T_\ka\phi_{ab}(x) \= \phi_{ab}(x)\ -\ \ka\int\!\diff^4y\ \Box^{-1}(x{-}y)\,(t_1\phi)_{ab}(y)\ +\ O(\ka^2) \ +\ O(\hbar)\ ,
\end{equation}
where $t_1\phi$ is quadratic in~$\phi_{\cdot\cdot}$ and of second order in derivatives.
With on-shell supersymmetry, we drop the auxiliary fields and only need to consider
\begin{equation}
\Delt^{\rm I} \= \tfrac{1}{2\ka}\,e\,\ga^{\mu\nu}D_\mu\psi_\nu
\= \tfrac1{2\ka}\,e\,e^\mu_{\ c}\,e^\nu_{\ d}\,\ga^{cd}\,(\pa_\mu+\tfrac14\om_{\mu ab}\ga^{ab})\,\psi_\nu
\end{equation}
of \eqref{M123} to order~$\ka$ so as to compute $\pa_\ka\Delt^{\rm inv}$ to leading order. 
To this end we expand (using $\approx$ to discard $O(\ka^3)$ contributions)
\begin{equation}
\begin{aligned}
e^a_\mu &\= \de^a_\mu + \ka\,\phi^a_\mu \ ,\qquad
e^\mu_a\ \approx\ \de^\mu_a-\ka\,\phi^\mu_a+\ka^2\phi{\cdot}\phi^\mu_a\ ,\qquad
e\ \approx\ 1+\ka\,\phi+\tfrac12\ka^2(\phi^2-\phi{:}\phi)\ ,\\[4pt]
\om_{\mu ab} &\ \approx\ \ka\,\bigl(\pa_b\phi_{a\mu}-\pa_a\phi_{b\mu}\bigr)\ +\ 
\tfrac14\ka^2\bigl(\bar\psi_\mu\ga_a\psi_b-\bar\psi_\mu\ga_b\psi_a+\bar\psi_a\ga_\mu\psi_b\bigr) \\
&\ +\ \tfrac12\ka^2\bigl(
\phi^\nu_a(2\pa_\nu\phi_{b\mu}-\pa_\mu\phi_{\nu b}-\pa_b\phi_{\mu\nu})-
\phi^\nu_b(2\pa_\nu\phi_{a\mu}-\pa_\mu\phi_{\nu a}-\pa_a\phi_{\mu\nu})-
\phi^\nu_\mu(\pa_a\phi_{b\nu}-\pa_b\phi_{a\nu})\bigr)\ ,
\end{aligned}
\end{equation}
where we abbreviated $\phi{\cdot}\phi^{\mu\nu}=\phi^\mu_a\phi^{a\nu}$ and $\phi{:}\phi=\phi^{ab}\phi_{ab}$
in addition to~$\phi=\phi^a_a$.
We will ignore the $\bar\psi\ga\psi$ contribution to the spin connection as a quantum correction to a Nicolai map.
Putting everything together, repeatedly partially integrating (since $\pa_\ka\Delt^{\rm I}$ is to be integrated over)
and making use of the gauge condition $\pa_b\phi^b_a=\tfrac12\pa_a\phi$, we finally arrive at
\begin{equation} \label{Mkappa}
\begin{aligned}
\pa_\ka\Delt^{\rm I} &\ \approx\ \bigl\{
\tfrac{5}{32}\phi^2 \eta^{\mu\nu} - \tfrac1{8}\phi{:}\phi\,\eta^{\mu\nu} - 
\tfrac38\phi\;\phi^{\mu\nu} + \tfrac14\phi{\cdot}\phi^{\mu\nu} +
\tfrac18\phi^2\ga^{\mu\nu} - \tfrac18\phi{:}\phi\,\ga^{\mu\nu} \\[2pt]
&\quad\ \ - \tfrac14\phi\;\phi^\mu_a\,\ga^{a\nu} + \tfrac14\phi\;\phi^\nu_a\,\ga^{a\mu} + 
\tfrac14\phi{\cdot}\phi^\mu_a\,\ga^{a\nu} - \tfrac14\phi{\cdot}\phi^\nu_a\,\ga^{a\mu} + 
\tfrac14\phi^\mu_a\phi^\nu_b\,\ga^{ab} \bigr\}\,\pa_\mu\psi_\nu\ .
\end{aligned}
\end{equation}
Unexpectedly, all derivatives could be moved onto the gravitino.
All possible terms of the form `$\phi\phi\,\pa\psi$' appear in~\eqref{Mkappa}.
Combining this with the free gravitino propagator~\eqref{psiprop} in the `Feynman' gauge~$\xi{=}\tfrac12$ 
and performing the spinor traces provides a leading-order classical flow operator
\begin{align}
R^{\rm inv}_{\ka=0} &\= \tfrac{2\im}{\hbar} \int\!\diff^4y\int\!\diff^4x\ \bigl\langle\,\pa_\ka\Delt^{\rm I}_\al(y)\ 
(-\tfrac12\bar\psi_\rho\ga^e)_\al(x)\,\bigr\rangle_{\ka=0}\,\tfrac{\de}{\de\phi^e_{\ \rho}(x)} \nn\\
&\= \int\!\diff^4y\int\!\diff^4x\ \bigl\{
\tfrac{9}{16}\phi^2\de^e_\rho\,\Box 
-\tfrac14\phi{:}\phi\,\de^e_\rho\,\Box 
-\tfrac54\phi\,\phi^\mu_\nu\,\de^e_\rho\,\pa_\mu\pa^\nu
+\phi{\cdot}\phi^\mu_\nu\,\de^e_\rho\,\pa_\mu\pa^\nu
-\phi\,\phi^e_\rho\,\Box
+\phi{\cdot}\phi^e_\rho\,\Box \\
&\qquad\qquad\qquad\qquad
+\phi^\mu_\nu\phi^e_\rho\,\pa_\mu\pa^\nu
-\phi^\mu_\rho\phi^e_\nu\,\pa_\mu\pa^\nu
+\phi\,\phi^\mu_\rho\,\pa_\mu\pa^e
-\phi{\cdot}\phi^\mu_\rho\,\pa_\mu\pa^e \bigr\}(y)\,
\Box^{-1}(y{-}x)\,\tfrac{\de}{\de\phi^e_{\ \rho}(x)} \nn
\end{align}
and therewith a potential Nicolai map~\eqref{mapleading} with
\begin{equation} \label{t1potential}
\begin{aligned}
(t_1\phi)_{ab} &\= \bigl(
-\tfrac{9}{16}\Box\,\phi^2
+\tfrac14\Box\,\phi{:}\phi
+\tfrac54\pa^c\pa^d\,\phi\,\phi_{cd}
-\pa^c\pa^d\,\phi{\cdot}\phi_{cd}
\bigr)\,\eta_{ab} 
+\Box\,\phi\,\phi_{ab}
-\Box\,\phi{\cdot}\phi_{ab} \\[2pt]
&\qquad
-\pa^c\pa^d\phi_{cd}\phi_{ab}
+\pa^c\pa^d\,\phi_{ac}\phi_{bd}
-\pa_{(a}\pa^c\phi\,\phi_{b)c}
+\pa_{(a}\pa^c\phi{\cdot}\phi_{b)c} \ ,
\end{aligned}
\end{equation}
where the partial derivatives here act on everything on their right.

Unfortunately, this candidate map does not satisfy the free-action condition.
Relaxing the coefficients to
\begin{align} \label{t1ansatz}
(t_1\phi)_{ab} &\= \bigl(
\la_1\,\Box\,\phi^2
+\la_2\,\Box\,\phi{:}\phi
+\la_3\,\pa^c\pa^d\,\phi\,\phi_{cd}
+\la_4\,\pa^c\pa^d\,\phi{\cdot}\phi_{cd}
\bigr)\,\eta_{ab}
+\la_5\,\Box\,\phi\,\phi_{ab}
+\la_6\,\Box\,\phi{\cdot}\phi_{ab} \\[2pt]
&\quad\
+\la_7\,\pa^c\pa^d\phi_{cd}\phi_{ab}
+\la_8\,\pa^c\pa^d\phi_{ac}\phi_{bd}
+\la_9\,\pa_{(a}\pa^c\phi\,\phi_{b)c}
+\la_{10}\,\pa_{(a}\pa^c\phi{\cdot}\phi_{b)c} 
+\la_{11}\,\pa_a\pa_b\,\phi^2
+\la_{12}\,\pa_a\pa_b\,\phi{:}\phi\ ,\nn
\end{align}
where two further possible index structures have been added,
gives us a more general ansatz.
In our gauge the Einstein--Hilbert lagrangian 
\begin{equation} 
\Lagr_{\textrm{\tiny EH}} \= \tfrac1{2\ka}\,e\,R \= \Lagr_2\ +\ \ka\,\Lagr_3\ +\ \ka^2\Lagr_4\ +\ O(\ka^3)
\end{equation}
has the leading parts~\cite{Nakajima:2025gwk, GoSa}\footnote{The ungauge-fixed cubic Lagrangian in~\cite{GoSa} contains 13 distinct terms. We suspect that twelve of them may be affected by a sign issue. This can be cross-checked by comparing it with the Lagrangian given in~\cite{Nakajima:2025gwk}.}
\begin{equation} \label{L2}
\Lagr_2 \= \tfrac12\,\phi^{ab}\,\Box\,\phi_{ab} - \tfrac14\,\phi\,\Box\,\phi
\end{equation}
and
\begin{equation} \label{L3}
\begin{aligned}
\Lagr_3 = & - \tfrac{1}{4} \phi^2 \Box \phi +
\tfrac{1}{4}\phi{:}\phi\,\Box\phi
+\tfrac12\,\phi\,\phi_{cd}\pa^c\pa^d\phi 
-\phi{\cdot}\phi_{cd}\pa^c\pa^d\phi
\\[2pt] &
+\tfrac12\,\phi\, \phi_{ab}\Box\phi^{ab} 
-\tfrac12\,\phi{\cdot}\phi_{ab}\Box\phi^{ab}
-\phi_{ab}\phi_{cd}\pa^c\pa^d\phi^{ab}
+2\,\phi_{ac}\phi_{bd}\pa^c\pa^d\phi^{ab}
\end{aligned}
\end{equation}
with only 8 of 12 possible structures showing up in~$\Lagr_3$.
Inserting~\eqref{mapleading} with~\eqref{t1ansatz} into  
\begin{equation}
\int\!\diff^4x\ \Lagr_2(T^{(0)}_\ka\phi_{..}) \ \buildrel{!}\over{=}\ 
\int\!\diff^4x\ \Lagr_2(\phi_{..})\ -\ \ka\int\!\diff^4x\ 
\bigl\{\phi^{ab} (t_1\phi)_{ab} - \tfrac12\,\phi\,(t_1\phi) \bigr\}\ +\ O(\ka^2)
\end{equation}
and matching with $\Lagr_3$ from~\eqref{L3} up to total derivatives yields the conditions
\begin{equation} \label{laconditions}
\bigl( \la_1,\,\la_2,\,\la_3,\,\la_4,\,\la_5,\,\la_6,\,\la_7,\,\la_8 \bigr)
\ \buildrel{!}\over{=}\ \bigl( 0,\, 0,\,0,\,0,\, -\tfrac12,\,\tfrac12,\,1,\,-2 \bigr)
\end{equation}
where the last four coefficients $\la_9,\ldots,\la_{12}$ remain arbitrary. 
Therefore, the free-action condition to leading order in~$\ka$ reduces the ansatz~\eqref{t1ansatz} to 
minimally 4 and maximally 8 terms.
Comparing with~\eqref{t1potential} we see that $\la_7{+}\la_8=0$ contradicts~\eqref{laconditions},
hence our candidate map with~\eqref{t1potential} obtained from $\pa_\ka\Delt^{\rm I}$ in~\eqref{Mkappa} is ruled out.
The technical reason is that
\begin{equation}
\de_\alpha \pa_\ka \Delt^{\rm I}_\al\!\mid_{\ka=0}\ \not\propto\ \Lagr_3
\end{equation}
as was the case for super-Yang--Mills. Replacing in~\eqref{Mkappa} $\psi_\nu$ by its supervariation and
taking the spinor trace, the first 4~terms do not contribute, and the 7 remaining terms generate 8 structures,
which all appear in~\eqref{L3}. But matching yields an overdetermined system
(8 linear equations for 7 relevant parameters in~\eqref{Mkappa}), and indeed the {\it single\/} last term in~\eqref{Mkappa}
produces the {\it two\/} last terms in~\eqref{L3} but with equal and opposite coefficients.
As a consequence, the on-shell construction of a (partial) Nicolai map fails for supergravity,
in contrast to super-Yang--Mills theory.

Of course, one may try to set up a map order by order in~$\ka$ by `brute force' via a general ansatz 
and imposing the Nicolai-map conditions (see~\eqref{matching01}) at each order.
At leading order, a minimal such map was found above, combining~\eqref{mapleading}, \eqref{t1ansatz} and~\eqref{laconditions},
\begin{equation} \label{minimalT}
\begin{aligned}
T_\ka\phi_{ab}(x) \=& \phi_{ab}(x)\ -\ \ka\int\!\diff^4y\ \Box^{-1}(x{-}y) \\
&\times\bigl\{
-\tfrac12\,\Box\,\phi\,\phi_{ab}
+\tfrac12\,\Box\,\phi{\cdot}\phi_{ab}
+\,\pa^c\pa^d\phi_{cd}\phi_{ab}
-2\,\pa^c\pa^d\phi_{ac}\phi_{bd} \bigr\}(y)\  +\ O(\ka^2)\ +\ O(\hbar) \ ,
\end{aligned}
\end{equation}
where the terms with $\la_9,\ldots,\la_{12}$ in~\eqref{t1ansatz} may be added at will.
To constrain the remaining $\lambda$~coefficients, we have to analyze to order~$\ka$ the quantum parts of the 
defining hierarchy of Nicolai-map conditions~\eqref{matchingnew}, the leading order~$\hbar$ of which is 
the determinant-matching condition spelled out in~\eqref{matching01}. Its right-hand side is given by the 
(log of the) fermion determinant, obtained by integrating out the gravitino in~\eqref{sugraaction} after dropping 
the four-fermi terms. Expanding as usual the determinant around the free one, to order~$\ka$ one extracts the
one-gravitino-loop one-point function from the $\bar\psi\phi\psi$ part of the action,
\begin{equation}
\begin{aligned}
S_\ka^{(1)}[\phi] &\= S_0^{(1)}\ +\ 
\Tr\,\langle \tfrac{e}{2\ka^2}\,R-\tfrac{e}{2}\,\bar\psi_\mu\ga^{\mu\nu\rho}D_\nu\psi_\rho\rangle|_{\bar\psi\phi\psi}
\ +\ O(\ka^2) \\
&\= S_0^{(1)}\ +\ 0\,\ka\,\delta^{(4)}(0) \int\!\diff^4x\ \phi(x) \ +\ O(\ka^2) \ ,
\end{aligned}
\end{equation}
where the formal delta-function singularity arises from a short-circuited free propagator.
The contributions from the Einstein-Hilbert term and the Rarita-Schwinger term cancel separately, 
leaving us with a vanishing $O(\ka)$ contribution to the fermion determinant.
On the left-hand side of~\eqref{matching01} appears the (log of the) Jacobian determinant for the classical map~$T_\ka^{(0)}$.
Computing the Jacobian of \eqref{mapleading} with~\eqref{t1ansatz} and performing the trace we arrive at
\begin{align}
- \im\,\Tr\ln\sfrac{\delta T_\ka^{(0)}\!\phi}{\delta\phi} &\= 
-\ka\,\bigl( 11\la_5+5\la_6+\sfrac{11}{2}\la_7+\sfrac{5}{2}\la_8+3\la_9+\sfrac{5}{2}\la_{10}+2\la_{11}+\la_{12}\bigr)\,
\delta^{(4)}(0) \int\!\diff^4x\ \phi(x) \ +\ O(\ka^2) \nn \\
&\= \ka\,\bigl( \tfrac52 - 3\la_9-\sfrac{5}{2}\la_{10}-2\la_{11}-\la_{12}\bigr)\,
\delta^{(4)}(0) \int\!\diff^4x\ \phi(x) \ +\ O(\ka^2)\ ,
\end{align}
where we inserted~\eqref{laconditions}.
To match the vanishing fermion determinant contribution, it seems that some of the (up to now free) coefficients
$\la_9,\ldots,\la_{12}$ need to be nonzero. For the full linear constraint on them, however, also the first term on the
left-hand side of~\eqref{matching01} is required, namely the leading quantum correction to the Nicolai map inserted
into the free bosonic action. The functional form of its $O(\ka)$ contribution is of the same (singular) form as the other 
terms in the determinant matching condition, but its coefficient will be subject to an extended ansatz for the map,
not fixed at this stage of the computation. Hence, our minimal map~\eqref{minimalT} may still respect also the $O(\hbar)$ 
Nicolai-map condition, but is more likely to need an enhancement by $\la_9,\ldots,\la_{12}$ terms 
(with a linear constraint on the coefficients). We expect the remaining ambiguity in the choice of the $\la$~parameters
to be eliminated only at order~$\ka^2$, where a larger variety of functional forms appear in the Nicolai-map conditions.

By allowing the two derivatives in each term of~$t_1\phi_{..}$ to also act separately on the $\phi$ factors,
a most general ansatz for the leading-order map may be written. It contains 21~terms and is presented in the Appendix
with its free-action constraints.

\section{Conclusions}\label{sec:outro}

\noindent
We encountered three obstacles in our attempt to construct a Nicolai map for minimal off-shell supergravity in four dimensions.
First, due to the superspace action being a superspace {\it density\/}, the off-shell supersymmetric lagrangian
cannot completely be written as a supervariation (even for rigid transformations).
This leads to a multiplicative term in the flow equation, which may be taken into account with a {\it partial\/} Nicolai map.
Second, supersymmetry now being part of the gauge invariance no longer (graded) commutes with the Slavnov (or BRST) variations
employed in the gauge-fixing procedure. As a consequence, we can use the BRST Ward identity only at the expense of another
multiplicative contribution to the flow equation.
Third, the rescaling trick required in the off-shell formalism in order to commute the supervariation and the derivative
with respect to the gravitational coupling fails, because the potentially singular part of the rescaled flow operator does
not entirely cancel with the functional Euler operator (as it does for super-Yang--Mills). Therefore, a perturbative
expansion of a Nicolai map appears to be obstructed in the off-shell formalism.
The mismatch is proportional to the trace of the metric fluctuation, which arises from the perturbative expansion of the
metric (or vierbein) determinant. 
This suggests that perhaps a unimodular version of supergravity~\cite{AMSG,BNPZ} may do better in this regard.

Finally, we applied the `trial-and-error' construction employing only on-shell supersymmetry, which is successful in the
super-Yang--Mills case, hoping at least for a {\it partial\/} Nicolai map.
However, the most general ansatz for $\pa_\ka\Delt$ does not correctly yield $\pa_\ka\Lagr_{\textrm{\tiny EH}}$ 
to the leading order upon a supervariation, ruling out also the on-shell construction. 
Nevertheless, relaxing the coefficients of the map ansatz so that they need not be obtained from a supervariation, 
we arrived at a four-parameter family of first-order Nicolai maps passing the free-action test.
More stringent tests await at the second order in the gravitational coupling and at order~$\hbar$,
a tedious but straightforward task beyond the scope of this work.

There are several ways in which the work presented here can be further expanded or generalized. First, pushing the `brute-force' 
ansatz to the second order and the quantum level and verifying the determinant-matching condition will be crucial.
Second, unimodular supergravity may overcome the third obstacle and allow for a proper perturbative off-shell flow operator.
Third, the consequences of a partial map should be investigated further, as they are relevant also for super-Yang--Mills
outside its critical dimensions.

\subsection*{Acknowledgments}
We thank Lorenzo Casarin for discussions.

\appendix
\section{General first-order ansatz}\label{sec:apx:onshell}

\noindent
The ansatz~\eqref{t1ansatz} of the form `$\pa\pa(\phi\phi)$' is somewhat special, motivated by the fact that in~\eqref{Mkappa} the derivatives could all be moved onto~$\psi$. A most general ansatz comprises all independent terms
of types `$\phi\pa\pa\phi$' and `$\pa\phi\pa\phi$'.
The special ansatz~\eqref{t1ansatz} can also be put in this form via
\begin{equation}
\tfrac12\int \Box^{-1}\,\pa\,\pa\,(\phi\,\phi) \= \int \Box^{-1}\,\bigl\{\phi\,\pa\,\pa\,\phi + \pa\,\phi\,\pa\,\phi\bigr\}\ .
\end{equation}
In total we have 11 plus 10 terms in the general ansatz,
\begin{equation} \label{t1Ansatz}
\begin{aligned}
(t_1\phi)_{ab} \ &=\
\la_{1a} \de_{ab} \phi \Box \phi + 
\la_{1b} \de_{ab} \phi^{cd} \pa_c \pa_d \phi +
\la_2 \de_{ab} \phi^{cd} \Box \phi_{cd} +
\la_3 \phi^{cd} \pa_a\pa_b \phi_{cd} +
\la_4 \phi^c_{(a} \Box \phi_{b)c} \\ &\ +
\la_{5a} \phi_{ab} \Box \phi +
\la_{5b} \phi \Box \phi_{ab} +
\la_{5c} \phi^{cd} \pa_c\pa_d \phi_{ab} +
\la_{6a} \phi \pa_a\pa_b \phi +
\la_{6b} \phi_{c(a} \pa^c\pa_{b)} \phi +
\la_{6c} \phi^{cd} \pa_d \pa_{(a} \phi_{b)c} \\ &\ +
\mu_{1a} \de_{ab} \pa^c \phi \pa_c \phi +
\mu_{1b} \de_{ab} \pa^c \phi^{de} \pa_e \phi_{cd} +
\mu_2 \de_{ab} \pa^c \phi^{de} \pa_c \phi_{de} +
\mu_3 \pa_a \phi^{cd} \pa_b \phi_{cd} +
\mu_4 \pa^d \phi^c_a \pa_d \phi_{bc} \\ &\ +
\mu_5 \pa^c \phi_{ab} \pa_c \phi +
\mu_{6a} \pa_a \phi \pa_b \phi +
\mu_{6b} \pa_{(a} \phi_{b)c} \pa^c \phi +
\mu_{6c} \pa^d \phi_{ac} \pa^c \phi_{bd} +
\mu_{6d} \pa^d \phi^c_{(a} \pa_{b)} \phi_{cd} \ ,
\end{aligned}
\end{equation}
where we labelled the coefficients such that $\la_{i*}$ and $\mu_{i*}$ terms are related by partial integration and~\eqref{landau}.
For the trace of~$\phi$, the ansatz simplifies to
\begin{equation} \label{t1trace}
\begin{aligned}
(t_1\phi)_a^a \ &=\
(4\la_{1a}{+}\la_{5a}{+}\la_{5b}{+}\la_{6a})\,\phi \Box \phi +
(4\la_2{+}\la_3{+}\la_4)\,\phi^{de} \Box \phi_{de} +
(4\la_{1b}{+}\la_{5c}{+}\la_{6b}{+}\tfrac12\la_{6c})\,\phi^{de} \pa_d\pa_e \phi  \\ &\ +
(4\mu_{1a}{+}\mu_5{+}\mu_{6a}{+}\tfrac12\mu_{6b})\,\pa^c \phi \pa_c \phi +
(4\mu_2{+}\mu_3{+}\mu_4)\,\pa^c \phi^{de} \pa_c \phi_{de} +
(4\mu_{1b}{+}\mu_{6c}{+}\mu_{6d})\,\pa^c \phi^{de} \pa_d \phi_{ce}\ .
\end{aligned}
\end{equation}

Inserting the candidate map with~\eqref{t1Ansatz} into the free action, one obtains at order~$\ka$ the expression
\begin{equation}
\begin{aligned}
& -\int\!\diff^4x\ \bigl\{\phi^{ab} (t_1\phi)_{ab} - \tfrac12\,\phi\,(t_1\phi) \bigr\}  
\\ \= &  \tfrac12\int\!\diff^4x\  \Bigl\{  \\ & 
+\big[- 2 \mu_{6a} + \la_{6b} - \tfrac12 \mu_{6c}- (2\la_{1b}{+}\la_{5c}{+}\la_{6b}{+}\tfrac12\la_{6c}  - 2
\la_{6a}) + \tfrac12 (2\mu_{1b}{+}\mu_{6c}{+}\mu_{6d})\big]  \phi^{ab} \pa_a \phi \pa_b \phi \\ &
+\big[- 2 \mu_3  + 2 (\la_{5c} + \la_3)\big] \phi^{ab} \pa_a \phi^{cd} \pa_b \phi_{cd} \\ &
+\big[- 2 \mu_{6b}  + 2 \la_{6b}  +
\la_{6c}- \mu_{6c} - (2\mu_{1b}{+}\mu_{6c}{+}\mu_{6d})\big] \phi^{ab} \pa_{a} \phi_{bc} \pa^c \phi \\ &
+\big[- 2 \mu_{6d} + 2 \la_{6c} + 2 \mu_{6c}\big] \phi^{ab} \pa^d \phi^c_{a} \pa_{b} \phi_{cd} \\ &
+\big[- 2 \mu_5  + 4 \la_{5a} + (\la_{5c} + \la_3) - (2\la_2{+}\la_3{+}\la_4-2 \la_{5b})\big] \phi^{ab} \pa^c \phi_{ab} \pa_c \phi \\ &
+\big[- 2 \mu_4  + 4 \la_4\big] \phi^{ab} \pa^d \phi^c_a \pa_d \phi_{bc}\\ &
+ \big[ (2\mu_{1a}{+}\mu_5{+}\mu_{6a}{+}\tfrac12\mu_{6b})-2 (2 \la_{1a}{+}\la_{5a}{+}\la_{5b}{+}\la_{6a} ) 
- \tfrac{1}{2}(2\la_{1b}{+}\la_{5c}{+}\la_{6b}{+}\tfrac12\la_{6c}  - 2 \la_{6a}) \\ 
& \qquad + \tfrac14 (2\mu_{1b}{+}\mu_{6c}{+}\mu_{6d})\big] \,\phi\,\pa^c \phi \pa_c \phi \\ &
+ \big[(2\mu_2{+}\mu_3{+}\mu_4) - (2\la_2{+}\la_3{+}\la_4-2 \la_{5b})\big] \,\phi\,\pa^c \phi^{de} \pa_c \phi_{de} \Bigr\} \ ,
\end{aligned}
\end{equation}
to be matched with $\int\Lagr_3$, which in this form reads~\cite{Nakajima:2025gwk, GoSa}
\begin{equation}
\begin{aligned}
\int\!\diff^4x\ \Bigl\{
& 0\,\phi^{ab} \pa_a \phi \pa_b \phi + \phi^{ab} \pa_a \phi^{cd} \pa_b \phi_{cd} + 0\,\phi^{ab} \pa_{a} \phi_{bc} \pa^c \phi - 2\,\phi^{ab} \pa^d \phi^c_{a} \pa_{b} \phi_{cd} \\ &
- \tfrac{1}{2}\,\phi^{ab} \pa^c \phi_{ab} \pa_c \phi + \,\phi^{ab} \pa^d \phi^c_a \pa_d \phi_{bc}+ \tfrac14\,\phi\,\pa^c \phi \pa_c \phi - \tfrac12 \,\phi\,\pa^c \phi^{de} \pa_c \phi_{de} \Bigr\} ~.
\end{aligned}
\end{equation}
This matching imposes the following constraints on the coefficients:
\begin{equation}
\begin{aligned}
0 &= 2\lambda_{6a} - 2\lambda_{1b} - \lambda_{5c} - \tfrac{1}{2}\lambda_{6c} 
     - 2\mu_{6a} - \tfrac{1}{2}\mu_{6c} + \tfrac{1}{2}\bigl(2\mu_{1b} + \mu_{6c} + \mu_{6d}\bigr), \\
1 &= \lambda_{3} + \lambda_{5c} - \mu_{3}, \\
0 &= 2\lambda_{6b} + \lambda_{6c} - 2\mu_{1b} - 2\mu_{6b} - 2\mu_{6c} - \mu_{6d}, \\
-2 &= \lambda_{6c} + \mu_{6c} - \mu_{6d}, \\
-1 &= -2\lambda_{2} - \lambda_{4} + 4\lambda_{5a} + 2\lambda_{5b} + \lambda_{5c} - 2\mu_{5}, \\
1 &= 2\lambda_{4} - \mu_{4}, \\
\tfrac{1}{2} &= -2\bigl(2\lambda_{1a} + \lambda_{5a} + \lambda_{6a} + \lambda_{5b}\bigr) 
     + \tfrac{1}{2}\bigl(2\lambda_{6a} - 2\lambda_{1b} - \lambda_{6b} - \lambda_{5c} - \tfrac{1}{2}\lambda_{6c}\bigr) \\
   &\quad + \mu_{5} + 2\mu_{1a} + \mu_{6a} + \tfrac{1}{2}\mu_{6b} 
     + \tfrac{1}{4}\bigl(2\mu_{1b} + \mu_{6c} + \mu_{6d}\bigr), \\
-1 &= -2\lambda_{2} - \lambda_{3} - \lambda_{4} + 2\lambda_{5b} + 2\mu_{2} + \mu_{3} + \mu_{4}~.
\end{aligned}
\end{equation}
This system of eight linear equations encodes the consistency conditions among the $\lambda$ and $\mu$~coefficients, 
ensuring that our most general ansatz reproduces the cubic part of the Einstein--Hilbert action. 
The general solution still depends on 13 independent parameters.
Taking these to be the 11 $\la$~parameters plus $\mu_{1b}$ and $\mu_{6c}$, the remaining 8~$\mu$ parameters may be fixed as follows,
\begin{equation}
\begin{aligned}
\mu_{1a} &= -\tfrac{1}{4} + \tfrac{1}{2}\lambda_{2} + \tfrac{1}{4}\lambda_{4} 
          + 2\lambda_{1a} + \lambda_{1b} + \tfrac{1}{2}\lambda_{5b} 
          + \tfrac{1}{4}\lambda_{5c} - \tfrac{1}{4}\mu_{1b}, \\
\mu_{2} &= \tfrac{1}{2} + \lambda_{2} - \tfrac{1}{2}\lambda_{4} - \lambda_{5b} - \tfrac{1}{2}\lambda_{5c}, \\
\mu_{3} &= -1 + \lambda_{3} + \lambda_{5c}, \\
\mu_{4} &= -1 + 2\lambda_{4}, \\
\mu_{5} &= \tfrac{1}{2} - \lambda_{2} - \tfrac{1}{2}\lambda_{4} 
          + 2\lambda_{5a} + \lambda_{5b} + \tfrac{1}{2}\lambda_{5c}, \\
\mu_{6a} &= \tfrac{1}{2} + \lambda_{6a} - \lambda_{1b} - \tfrac{1}{2}\lambda_{5c} 
          + \tfrac{1}{2}\mu_{1b} + \tfrac{1}{4}\mu_{6c}, \\
\mu_{6b} &= -1 + \lambda_{6b} - \mu_{1b} - \tfrac{3}{2}\mu_{6c}, \\
\mu_{6d} &= 2 + \lambda_{6c} + \mu_{6c}~.
\end{aligned}
\end{equation}

\newpage

%\section*{}

\end{document}